\documentclass[aps,showpacs,prl,twocolumn,groupedaddress,superscriptaddress,floatfix]{revtex4}
\usepackage{graphicx}
\usepackage{bm}

\newcommand{\gsim}{\raisebox{-0.7ex}{$\stackrel{\textstyle >}{\sim}$ }}

\def\si{^1 \hskip -0.02in S _0}
\def\siii{^3 \hskip -0.02in S _1}
\def\diii{^3 \hskip -0.04in D _1}

\begin{document}

\title {Nucleon-Nucleon Scattering From Fully-Dynamical Lattice QCD}

\author{\bf S.R.~Beane}
\affiliation{Department of Physics, University of New Hampshire, Durham, NH 03824-3568}
\affiliation{Jefferson Laboratory, 12000 Jefferson Avenue, Newport News, VA 23606}
\author{\bf P.F.~Bedaque}
\affiliation{Lawrence-Berkeley Laboratory, Berkeley, CA 94720}
\affiliation{Department of Physics, University of Maryland, College Park, MD 20742}
\author{\bf K.~Orginos}
\affiliation{Department of Physics, College of William and Mary, Williamsburg, VA 23187-8795}
\affiliation{Jefferson Laboratory, 12000 Jefferson Avenue, Newport News, VA 23606}
\author{\bf M.J.~Savage}
\affiliation{Department of Physics, University of Washington, Seattle, WA 98195-1560}
\collaboration{ NPLQCD Collaboration }
\noaffiliation

\begin{abstract}
We present results of the first fully-dynamical lattice QCD
determination of nucleon-nucleon scattering lengths in the $\si$
channel and $\siii-\diii$ coupled channels.  The calculations are
performed with domain-wall valence quarks on the MILC staggered
configurations with lattice spacing of $b=0.125~{\rm fm}$ in the
isospin-symmetric limit, and in the absence of electromagnetic
interactions.
\end{abstract}
\maketitle

One of the ultimate goals of nuclear physics is to compute the
properties and interactions of nuclei directly from Quantum
Chromodynamics (QCD), the underlying theory of the strong
interactions. Achievement of this goal would reveal how nuclear
processes depend upon the fundamental constants of nature, and would
enable the computation of strong-interaction processes of importance
in environments not attainable in the laboratory, such as in the
interior of neutron stars.  A lone, pioneering study of
nucleon-nucleon ($NN$) scattering with lattice QCD was performed more
than a decade ago by Fukugita~{\it et al}~\cite{Fukugita:1994ve}.
These calculations were quenched and 
at relatively large pion masses, $m_\pi\gsim 550~{\rm MeV}$.

Lattice QCD is presently the only known method of calculating
low-energy strong-interaction processes from QCD.  However, it is
unlikely that lattice QCD will ever be used to calculate the
properties or interactions of nuclei beyond the lightest few.
In order to compute the properties of larger nuclei,
lattice QCD calculations of the lightest nuclei will be performed, and
then matched to calculations of the larger nuclei using many-body
techniques such as Greens Function Monte-Carlo (GFMC),
e.g. Ref.~\cite{Pieper:2004qw}, or the No-Core Shell Model (NCSM),
e.g. Ref.~\cite{Forssen:2004dk}, including up to 
four-body (and possibly higher) interactions
consistent with chiral symmetry and
power-counting.  These many-body methods have had great
success in reproducing the properties of the light nuclei using
high-precision phenomenological potentials as input and recently
with the chiral potentials derived from effective field theory (EFT)
expansions.

In this letter we present results of the first fully-dynamical lattice
QCD calculation of the $NN$ scattering lengths in both the
$\si$-channel and $\siii-\diii$-coupled-channels at pion masses of
$m_\pi \sim 350~{\rm MeV}$ , $490~{\rm MeV}$ and $590~{\rm MeV}$ in
the isospin-limit.  Our lattice calculations are performed with
domain-wall~\cite{Kaplan:1992bt,Shamir:1993zy} valence quarks on the
$20^3\times 64$ MILC gauge-field configurations with a lattice spacing
of $b\sim 0.125~{\rm fm}$ and a spatial extent of $L\sim 2.5~{\rm
fm}$. The dependence of the $NN$ scattering lengths upon the
light-quark masses has been determined to various non-trivial orders
in the EFT expansion~\cite{Beane:2002vs,Beane:2002xf,Epelbaum:2002gb},
and is estimated to be valid up to $m_\pi\sim 350~{\rm MeV}$.  We use
the results of our lattice QCD calculation at the lightest pion mass
and the experimentally-determined scattering lengths at the physical
value of the pion mass to constrain the chiral dependence of the scattering lengths from
$m_\pi\sim 350~{\rm MeV}$ down to the chiral limit.

The $NN$ scattering lengths are determined by computing the energy shift of the
lowest-lying two-nucleon state in the finite lattice volume.  To
extract $p\cot\delta(p)$ ---where $\delta(p)$ is the phase shift--- the
magnitude of the center-of-mass momentum, $p$, is extracted from this
energy shift and inserted into L\"uscher's relation~\cite{Luscher:1990ux}
\begin{eqnarray}
p\cot\delta(p) \ =\ {1\over \pi L}\ \left( \sum_{\bf j}^{ |{\bf j}|<\Lambda}
{1\over |{\bf j}|^2-(\frac{p L}{2\pi})^2}\ -\  {4 \pi \Lambda} \right)
\ \ ,
\label{eq:energies}
\end{eqnarray}
which is valid below the inelastic threshold. The sum in eq.~(\ref{eq:energies})
is over all triplets of integers ${\bf j}$ such that $|{\bf j}| < \Lambda$ and the
limit $\Lambda\rightarrow\infty$ is implicit~\cite{Beane:2003da,Beane:2003yx}.
The s-wave scattering lengths in the $NN$ systems follow 
from the effective range expansion (ERE)
\begin{eqnarray}
 p\cot\delta(p) & = & -{1\over a}\ +\ {1\over 2}\;r\;p^2\ +\ ...
\ \ \ ,
\label{eq:pcotdelta}
\end{eqnarray}
where $a$ is the scattering length (using the nuclear physics
sign convention) and $r$ is the effective range which is typically of order the
range of the interaction, $\sim 1/m_\pi$.

Following the LHP collaboration (LHPC)~\cite{Edwards:2005kw,Renner:2004ck}, our computation is a
hybrid lattice QCD calculation using domain-wall valence quarks from a
smeared-source on three sets of $N_f=2+1$
asqtad-improved~\cite{Orginos:1999cr,Orginos:1998ue}
MILC configurations generated
with staggered sea quarks~\cite{Bernard:2001av}. In the generation of
the MILC configurations, the strange-quark mass was fixed near its
physical value, $b m_s = 0.050$, (where $b$ is the lattice spacing)
determined by the mass of hadrons containing strange quarks.  The two
light quarks in the three sets of configurations are degenerate
(isospin-symmetric), with masses $b m_l=0.010, 0.020$ and $0.030$.
Some of the domain-wall valence propagators were previously generated
by LHPC on each of these sets of lattices.  The domain-wall height is
$m=1.7$ and the extent of the extra dimension is $L_5=16$.  The
parameters used to generate the light-quark propagators have been
``tuned'' so that the mass of the pion computed with the domain-wall
propagators is equal (to few-percent precision) to that of the
lightest staggered pion computed with the same parameters as the gauge
configurations~\cite{Bernard:2001av}. The MILC lattices were  
HYP-blocked~\cite{Hasenfratz:2001hp} and Dirichlet boundary conditions
were used to reduce the time extent of the MILC lattices from 64 to 32
time-slices in order to save time in propagator generation.
Various parts of the lattice were employed to generate multiple sets
of propagators on each lattice.  We analyzed one set of correlation
functions on 564 lattices with $b m_l=0.030$, three sets of
correlation functions on 486 lattices with $b m_l=0.020$, and one set
of correlation functions on 658 lattices with $b m_l=0.010$.  The
lattice calculations were performed with the {\it Chroma} software
suite~\cite{Edwards:2004sx,sse2} on the high-performance computing
systems at the Jefferson Laboratory (JLab).

\begin{figure}[!ht]
\vskip 0.25in
\centerline{\includegraphics[width=2.6in]{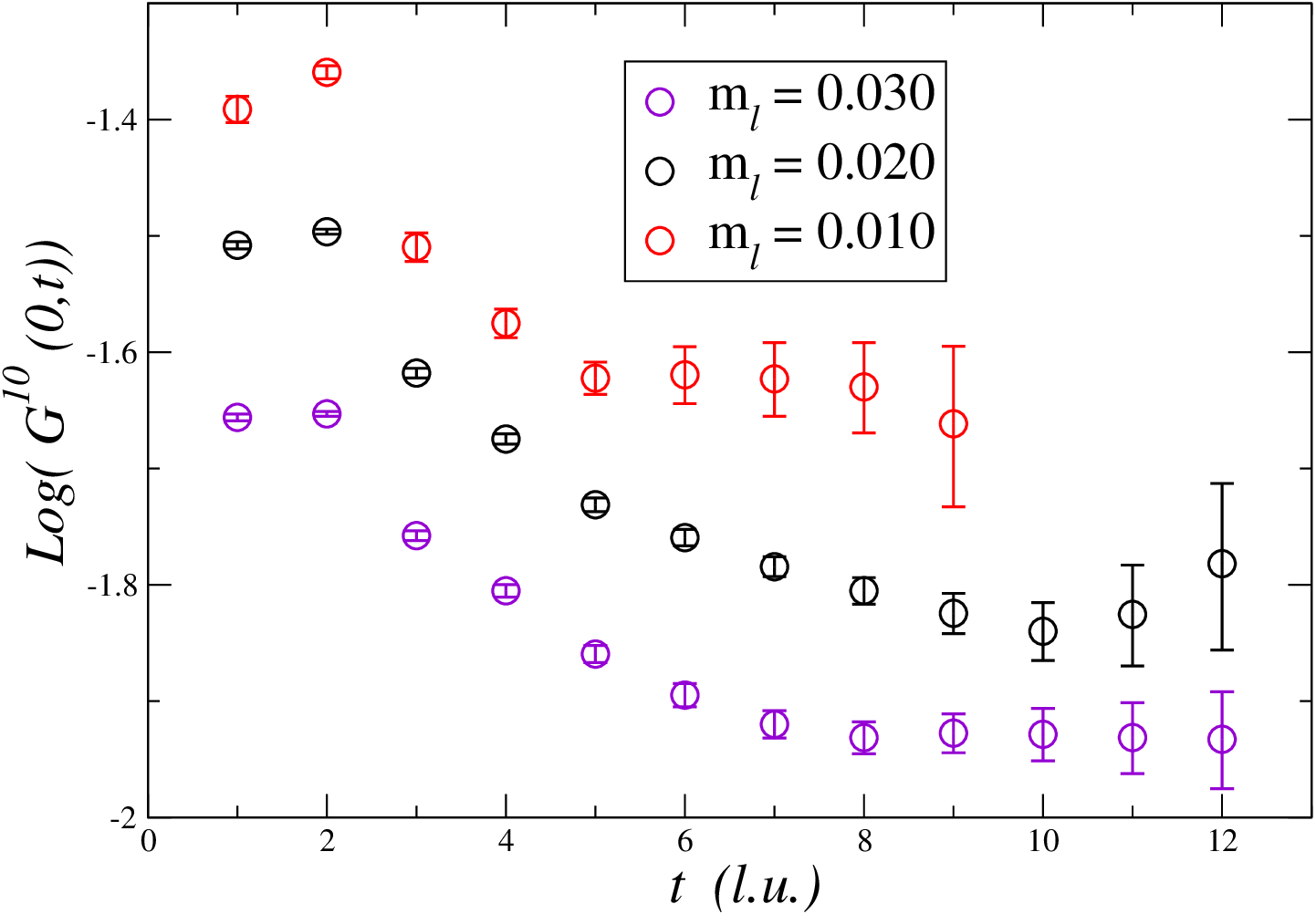}}
\noindent
\caption{
The logarithm of the ratio of correlation functions
in the $\si$ channel  ($G^{10}$) as a function of time-slice.
Each has been off-set vertically for display purposes.
}
\label{fig:logcorr_sing}
\end{figure}

\begin{figure}[!ht]
\vskip 0.15in
\centerline{\includegraphics[width=2.6in]{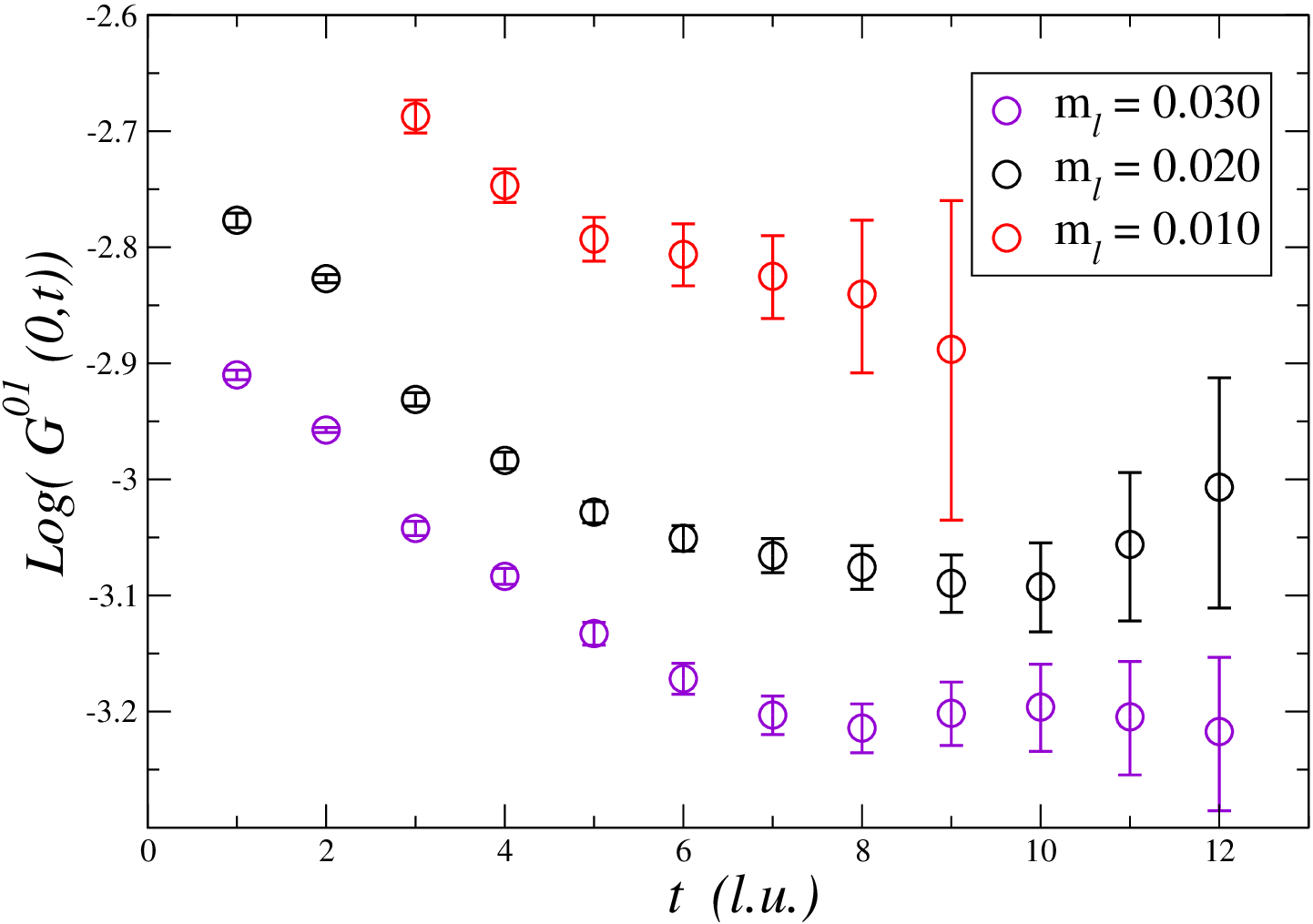}}
\noindent
\caption{
The logarithm of the ratio of correlation functions
in the $\siii-\diii$ coupled-channels ($G^{01}$)
as a function of time-slice.
Each has been off-set vertically for display purposes.
}
\label{fig:logcorr_trip}
\end{figure}

The contractions necessary to produce the required correlation functions
were performed in two stages.  First, three propagators were
contracted together at a sink with the quantum numbers of the nucleon,
and Fourier transformed on each time slice to produce a non-lattice
object with multiple Dirac and isospin indices (on the initial
time slice).  Second, two of these objects were contracted together to
produce the two-nucleon correlation functions as a function of time-slice.  
For an arbitrary nucleus (or bound and continuum nucleons), 
of atomic number $A$ and charge $Z$, there are $(A+Z)!\
(2A-Z)!$ contractions that must be performed to produce the nuclear
correlation function.  Therefore, in the $\si$ channel there are $48$
contractions, while in the $\siii-\diii$ coupled channels system there
are $36$.  The
cleanest quantity from which to extract the energy-difference between the
two-nucleon state(s) and the mass of two noninteracting nucleons was found to be
the ratio of correlation functions
\begin{eqnarray}
G^{IS}(t) & = & C_{NN}^{IS}(t)/\left( C_{N}(t)\right)^2
\ \ ,
\label{eq:ratio}
\end{eqnarray}
where $I$ denotes the isospin of the $NN$ system and $S$ denotes its spin.
The single-nucleon correlator is
\begin{eqnarray}
C_{N}(t) & = & \sum_{\bf x}
\langle N(t,{\bf x})\ N^\dagger(0, {\bf 0})
\rangle
\ \ \ ,
\label{N_correlator} 
\end{eqnarray}
and the two-nucleon correlator that projects onto the s-wave state 
in the continuum limit is
\begin{eqnarray}
& & C_{NN}^{IS}(t) \ =\  
X_{\alpha\beta\sigma\rho}^{ijkl}
\nonumber\\
& & \qquad
\sum_{\bf x , y}
\langle N^\alpha_i(t,{\bf x})N^\beta_j(t, {\bf y})
N^{\sigma\dagger}_k (0, {\bf 0})N^{\rho\dagger}_l(0, {\bf 0})
\rangle
\  , 
\label{NN_correlator} 
\end{eqnarray}
where $\alpha,\beta,\sigma,\rho$ are isospin-indices and $i,j,k,l$ are
Dirac-indices.
The tensor $X_{\alpha\beta\sigma\rho}^{ijkl}$ has elements that produce the
correct spin-isospin quantum numbers of two-nucleons in an s-wave.
The summation over ${\bf x}$ (and ${\bf y}$) corresponds to summing over all
the spatial lattice sites, thereby projecting onto 
the momentum ${\bf p}={\bf 0}$ state.
The interpolating field for the proton is
$p_i(t,{\bf x}) = \epsilon_{abc} u^a_i(t, {\bf x}) \left( u^{b T}(t, {\bf x})
  C\gamma_5 d^c(t, {\bf x})\right)$, and similarly for the neutron.
The ratio of correlation functions that we obtain in the$\si$ channel and the 
$\siii-\diii$ coupled channels,
at the three different pion
masses, are shown in fig.~\ref{fig:logcorr_sing} and
fig.~\ref{fig:logcorr_trip}, 
respectively.

The results of our calculations are shown in Table~\ref{table:scattlengths}.
The scale is set via the quark-mass dependence of $f_\pi$, which gives
$b=0.127 \pm 0.001~{\rm fm}$~\cite{Beane:2005rj} (consistent
with the Sommer scale-setting procedure used by MILC~\cite{{Bernard:2001av}}).
\begin{table}[ht]
\begin{tabular}{|c|c|c|}
\hline
\quad $m_\pi\  (MeV)$\quad  
&\quad $a{(\si)}\  (fm)$\quad 
&\quad $a{(\siii)}\  (fm)$\quad  \\
       \hline
 \quad $353.7 \pm 2.1$\quad     & \quad $0.63\pm 0.50$ (5-10)\quad        & \quad $0.63\pm 0.74$ (5-9)\quad \\
 \quad $492.5 \pm 1.1$\quad      & \quad $0.65 \pm 0.18$ (6-9)\quad       & \quad $0.41\pm 0.28$ (6-9)\quad \\
 \quad $593.0 \pm 1.6$\quad     & \quad $0.0\pm 0.5$ (7-12)\quad        & \quad $-0.2\pm 1.3$ (7-12)\quad \\
  \hline
  \end{tabular}
\vskip 0.1in
  \caption{Scattering lengths in the $\si$ channel and in the 
    $\siii-\diii$ coupled channels.  The
    uncertainty is statistical and the fitting ranges are in parentheses. 
    There is a systematic error of
    $\sim 0.1~{\rm fm}$ on each scattering length associated with the truncation
    of the effective range expansion; i.e. the numbers exhibited are for
    $-1/p\cot\delta$ at the measured energy-splitting.}
\vskip 0.1in
\label{table:scattlengths}
\end{table}
At the pion masses used in these calculations the $NN$ scattering
lengths are found to be of natural size in both channels, and are much
smaller than the $L\sim 2.5~{\rm fm}$ lattice spatial extent.  It is
noteworthy that our scattering lengths at the heaviest pion mass 
are not inconsistent with the lightest-mass quenched values of Ref.~\cite{Fukugita:1994ve}. However,
one should keep in mind the effects of quenching on the infrared properties of the
theory~\cite{Beane:2002nu}.

\begin{figure}[!ht]
\vskip 0.25in
\centerline{\includegraphics[width=2.8in]{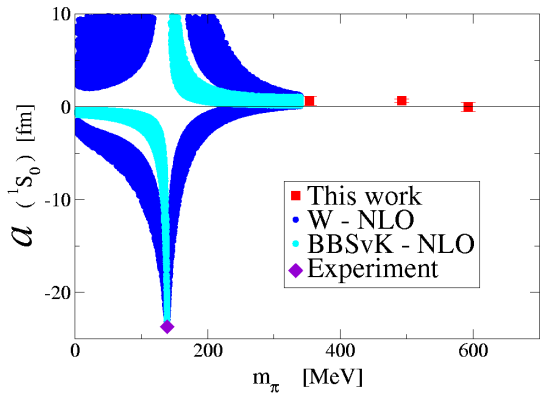}}
\noindent
\caption{Allowed regions for the scattering length in the $\si$ channel as a
  function of the pion mass. The experimental value of the scattering length 
  and NDA have been used to constrain the  extrapolation in both 
BBSvK~\protect\cite{Kaplan:1998we,Kaplan:1998tg,Beane:2001bc} and W~\protect\cite{Weinberg:1990rz,Weinberg:1991um,Ordonez:1995rz} 
power-countings at NLO.
}
\label{fig:ampi_sing}
\end{figure}

\begin{figure}[!ht]
\vskip 0.15in
\includegraphics[width=7.0cm]{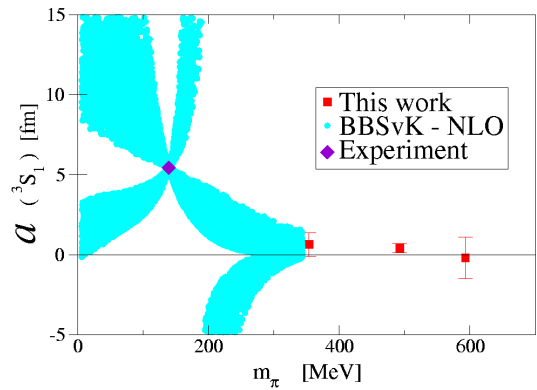}
\noindent
\caption{Allowed regions for the scattering length in the $\siii-\diii$ 
coupled-channels as a function of the pion mass. 
The experimental value of the scattering length and NDA have been used to constrain the 
extrapolation in BBSvK~\protect~\cite{Beane:2001bc} power-counting at NLO. (W counting
gives essentially identical results.)
}
\label{fig:ampi_trip}
\end{figure}

The lowest pion mass at which we have calculated is at the upper limit
of where we expect the EFT describing $NN$ interactions to be
valid~\cite{Weinberg:1990rz,Weinberg:1991um,Ordonez:1995rz,Kaplan:1998we,Kaplan:1998tg,Beane:2001bc}.
While some controversy remains regarding the details of the $NN$ EFT,
in our present analysis, we have constrained the chiral extrapolation
using BBSvK power-counting~\cite{Beane:2001bc} ($\equiv$KSW
power-counting~\cite{Kaplan:1998we,Kaplan:1998tg}) and W
power-counting~\cite{Weinberg:1990rz,Weinberg:1991um,Ordonez:1995rz}
in the $\si$-channel and BBSvK power-counting in the $\siii-\diii$
coupled channels.  The recent lattice QCD determinations of the
light-quark axial-matrix element in the nucleon by
LHPC~\cite{Edwards:2005ym} and its physical value are used to
constrain the chiral expansion of $g_A$. Our lattice calculations of
the nucleon mass and pion decay constant~\cite{Beane:2005rj} ---as
well as their physical values--- are used to constrain their
respective chiral expansions.  In addition to the quark-mass
dependence these three quantities contribute to the $NN$ systems,
there is dependence on the quark masses at next-to-leading order (NLO)
from pion exchange, and from local four-nucleon operators that
involve a single insertion of the light-quark mass matrix, described
by the ``$D_2$'' coefficients~\cite{Beane:2002vs,Beane:2002xf,Epelbaum:2002gb}.  The
results of this lattice QCD calculation constrain the range of allowed
values for the $D_2$'s, and consequently the scattering lengths in the
region between $m_\pi\sim 350~{\rm MeV}$ and the chiral limit, as
shown in fig.~\ref{fig:ampi_sing} and fig.~\ref{fig:ampi_trip}.  With
only one lattice point at the edge of the regime of applicability of
the EFT, a prediction for the scattering lengths at the physical pion
mass is not possible: the experimental values of the scattering
lengths are still required for an extrapolation to the chiral limit
and naive dimensional analysis (NDA) is still required to select only
those operator coefficients that are consistent with perturbation
theory. The regions plotted in the figures correspond to values
of $C_0$ -- the coefficient of the leading-order quark-mass independent
local operator -- and $D_2$ that fit the lattice
datum and the physical value, and are consistent with NDA; indeed we have $D_2(\Lambda)
m_\pi^2/C_0(\Lambda) \sim \pm 0.10$ in both channels (at physical
$m_\pi$), at a renormalization scale $\Lambda\sim 350~{\rm MeV}$.
In both channels the lightest lattice datum constrains the
chiral extrapolation to two distinct bands which are sensitive to both
the quark mass dependence of $g_A$ and the sign of the $D_2$
coefficient. 
As the lattice point used to constrain the EFT is at the upper limits
of applicability of the EFT, we expect non-negligible corrections to
these regions from higher orders in the EFT expansion.  It is clear
from fig.~\ref{fig:ampi_sing} and fig.~\ref{fig:ampi_trip} that even a
qualitative understanding of the chiral limit will require lattice calculations
at lighter quark masses.


Without the resources to perform similar lattice QCD calculations in different volumes, 
and observing that most energy-splitting are positive,
we have assumed that we have observed scattering states
in each case.
Calculations in larger volumes will be done in the future to verify the
expected power-law dependence upon volume that scattering states exhibit.
In addition to discriminating between bound and continuum
states, calculations in a larger volume would reduce the energy of the
lowest-lying continuum lattice states, and thus reduce the uncertainty
in the scattering length due to truncation of the ERE. 
Further improvement would result from measuring the energy of the
first excited state on the lattice, either with a single source or by
using the L\"uscher-Wolff~\cite{Luscher:1990ck} method.

There are many aspects of this calculation that should be refined in
the future.  The statistics should be improved by at least an order of
magnitude to have a precise extraction of the scattering lengths from
each of these lattices.  The lattice spacing effects in the present
calculation appear at $\sim {\cal O}(b^2)$ (or exponentially
suppressed ${\cal O}(b)$ effects), and are expected to be small.
However, the finite lattice spacing effects should be determined by
performing the same calculation on lattices with a finer lattice
spacing. While it would be useful to perform this calculation with
configurations generated from domain-wall quarks in the sea sector as
well, the resources to do so are not currently available to
us. However, a recent theoretical investigation~\cite{Chen:2005ab} of
the impact of using a mixed-action to compute $\pi\pi$
scattering~\cite{Beane:2005rj} has shown it to be small.  An analogous
theoretical investigation of mixed-action EFT for $NN$ scattering
which would allow a continuum extrapolation remains to be carried
through.  In addition to more precise lattice QCD calculations through
an increase in computing resources, formal developments are also
required.  In order to have a more precise chiral extrapolation,
calculations in the various relevant EFT's must be performed beyond
NLO.  Furthermore, it is clear that lattice calculations at lower pion
masses are essential for the extrapolation to the chiral limit, and
will ultimately allow for a ``prediction'' of the physical scattering
lengths.

To summarize, we have performed the first QCD calculations of
nucleon-nucleon scattering by using fully-dynamical, mixed action
lattice QCD.  This work opens up an unexplored area of nuclear
physics as it is now possible to perform lattice QCD
calculations of simple nuclear systems at pion masses within the range
of validity of the $NN$ EFT's.

\vskip 0.1in

\acknowledgments We would like to thank the high-performance computing
group at JLab. We would also like to thank LHPC for use of their
resources. This work was supported by the U.S.~Department of Energy
through the SciDAC project and through grants/contracts:
No.~DE-FG03-97ER4014 (MJS, NT@UW-06-01), No.~DF-FC02-94ER40818 (KO),
No.~DE-AC03-76SF00098 (PFB, LBNL-59535) and No.~DE-AC05-84ER40150
(SRB, KO, JLAB-THY-06-462), and by the National Science Foundation through grant
No.~PHY-0400231 (SRB, UNH-06-01).


\begin{thebibliography}{99}
\bibitem{Fukugita:1994ve}
  M.~Fukugita, Y.~Kuramashi, M.~Okawa, H.~Mino and A.~Ukawa,
  Phys.\ Rev.\ D {\bf 52}, 3003 (1995)
  [arXiv:hep-lat/9501024].

\bibitem{Pieper:2004qw}
  S.~C.~Pieper, R.~B.~Wiringa and J.~Carlson,
  Phys.\ Rev.\ C {\bf 70}, 054325 (2004)
  [arXiv:nucl-th/0409012].

\bibitem{Forssen:2004dk}
  C.~Forssen, P.~Navratil, W.~E.~Ormand and E.~Caurier,
  Phys.\ Rev.\ C {\bf 71}, 044312 (2005)
  [arXiv:nucl-th/0412049].

\bibitem{Kaplan:1992bt}
  D.~B.~Kaplan,
  Phys.\ Lett.\ B {\bf 288}, 342 (1992)
  [arXiv:hep-lat/9206013].

\bibitem{Shamir:1993zy}
  Y.~Shamir,
  Nucl.\ Phys.\ B {\bf 406}, 90 (1993)
  [arXiv:hep-lat/9303005].

\bibitem{Beane:2002vs}
  S.~R.~Beane and M.~J.~Savage,
  Nucl.\ Phys.\ A {\bf 713}, 148 (2003)
  [arXiv:hep-ph/0206113].

\bibitem{Beane:2002xf}
  S.~R.~Beane and M.~J.~Savage,
  Nucl.\ Phys.\ A {\bf 717}, 91 (2003)
  [arXiv:nucl-th/0208021].

\bibitem{Epelbaum:2002gb}
  E.~Epelbaum, U.~G.~Meissner and W.~Gloeckle,
  Nucl.\ Phys.\ A {\bf 714}, 535 (2003)
  [arXiv:nucl-th/0207089].

\bibitem{Luscher:1990ux}
  M.~L\"uscher,
  Nucl.\ Phys.\ B {\bf 354}, 531 (1991).

\bibitem{Beane:2003da}
  S.~R.~Beane, P.~F.~Bedaque, A.~Parre\~no and M.~J.~Savage,
  Phys.\ Lett.\ B {\bf 585}, 106 (2004)
  [arXiv:hep-lat/0312004].

\bibitem{Beane:2003yx}
  S.~R.~Beane, P.~F.~Bedaque, A.~Parre\~no and M.~J.~Savage,
  Nucl.\ Phys.\ A {\bf 747}, 55 (2005)
  [arXiv:nucl-th/0311027].

\bibitem{Edwards:2005kw}
  R.~G.~Edwards {\it et al.}  [LHPC Collaboration],
  PoS {\bf LAT2005}, 056 (2005)
  [arXiv:hep-lat/0509185].

\bibitem{Renner:2004ck}
  D.~B.~Renner {\it et al.}  [LHP Collaboration],
  Nucl.\ Phys.\ Proc.\ Suppl.\  {\bf 140}, 255 (2005)
  [arXiv:hep-lat/0409130].


\bibitem{Orginos:1999cr}
  K.~Orginos, D.~Toussaint and R.~L.~Sugar  [MILC Collaboration],
  Phys.\ Rev.\ D {\bf 60}, 054503 (1999)
  [arXiv:hep-lat/9903032].

\bibitem{Orginos:1998ue}
  K.~Orginos and D.~Toussaint  [MILC collaboration],
  Phys.\ Rev.\ D {\bf 59}, 014501 (1999)
  [arXiv:hep-lat/9805009].

\bibitem{Bernard:2001av}
  C.~W.~Bernard {\it et al.},
  Phys.\ Rev.\ D {\bf 64}, 054506 (2001)
  [arXiv:hep-lat/0104002],
  http://qcd.nersc.gov/.

\bibitem{Hasenfratz:2001hp}
  A.~Hasenfratz and F.~Knechtli,
  Phys.\ Rev.\ D {\bf 64}, 034504 (2001)
  [arXiv:hep-lat/0103029].

\bibitem{Edwards:2004sx}
  R.~G.~Edwards and B.~Joo  [SciDAC Collaboration],
  Nucl.\ Phys.\ Proc.\ Suppl.\  {\bf 140} (2005) 832
  [arXiv:hep-lat/0409003].

\bibitem{sse2}
C.~McClendon, 
Jlab preprint, JLAB-THY-01-29 (2001).

\bibitem{Beane:2005rj}
  S.~R.~Beane, P.~F.~Bedaque, K.~Orginos and M.~J.~Savage  [NPLQCD
                  Collaboration],
  [arXiv:hep-lat/0506013].

\bibitem{Beane:2002nu}
  S.~R.~Beane and M.~J.~Savage,
  Phys.\ Lett.\ B {\bf 535}, 177 (2002)
  [arXiv:hep-lat/0202013].

\bibitem{Weinberg:1990rz}
  S.~Weinberg,
  Phys.\ Lett.\ B {\bf 251}, 288 (1990).

\bibitem{Weinberg:1991um}
  S.~Weinberg,
  Nucl.\ Phys.\ B {\bf 363}, 3 (1991).

\bibitem{Ordonez:1995rz}
  C.~Ordo\~nez, L.~Ray and U.~van Kolck,
  Phys.\ Rev.\ C {\bf 53}, 2086 (1996)
  [arXiv:hep-ph/9511380].

\bibitem{Kaplan:1998we}
  D.~B.~Kaplan, M.~J.~Savage and M.~B.~Wise,
  Nucl.\ Phys.\ B {\bf 534}, 329 (1998)
  [arXiv:nucl-th/9802075].

\bibitem{Kaplan:1998tg}
  D.~B.~Kaplan, M.~J.~Savage and M.~B.~Wise,
  Phys.\ Lett.\ B {\bf 424}, 390 (1998)
  [arXiv:nucl-th/9801034].

\bibitem{Beane:2001bc}
  S.~R.~Beane, P.~F.~Bedaque, M.~J.~Savage and U.~van Kolck,
  Nucl.\ Phys.\ A {\bf 700}, 377 (2002)
  [arXiv:nucl-th/0104030].

\bibitem{Edwards:2005ym}
  R.~G.~Edwards {\it et al.}  [LHPC Collaboration],
  [arXiv:hep-lat/0510062].

\bibitem{Luscher:1990ck}
  M.~L\"uscher and U.~Wolff,
  Nucl.\ Phys.\ B {\bf 339}, 222 (1990).

\bibitem{Chen:2005ab}
  J.~W.~Chen {\it et al},
  [arXiv:hep-lat/0510024].




\end{thebibliography}
\end{document}